\documentstyle{elsart}

\begin{document}
\begin{frontmatter}

\title{
     SCINTILLATION PROPORTIONAL Xe COUNTER WITH WLS FIBER \\
     READOUT FOR LOW-ENERGY X-RAYS
\\}
\collab{
     D. Akimov$^1$ , S. Belogurov, A. Burenkov, D. Churakov,\\
     V. Kuzichev, V. Morgunov, G. Smirnov, V. Solovov\\}
\address{ Institute for Theoretical and Experimental Physics \\
  RU-117279  Moscow  Russia \\ }
\date{Jan. 25, 1997}
\thanks[CA]{Corresponding author. 
 Institute for Theoretical and Experimental Physics;
 B. Cheremushkinskaya, 25, Moscow, Russia, 117259
 Tel.: +7 095 125 9293   Fax:  +7 095 123 6584
 e-mail:Akimov\_D@vitep3.itep.ru }

\begin{abstract}
A gas Xe based scintillation proportional counter with cylindrical geometry
and wavelength shifting (WLS) fiber readout for X-rays of energy
$0.5\; - \; 100\; keV$ is proposed. With such a design large sizes and 
sensitive area of the counter with a fairly well uniformity is possible.
The counter could be used for "dark matter" search and neutrino magnetic moment 
measurement and for detection of small amounts or traces of radioactive 
elements in substances or environment.
\end{abstract}
\end{frontmatter}

\section*{Introduction}

     The electroluminescent (scintillation proportional) counters with a
single anode wire and a PMT readout having a "pill-box" and cylinder geometry
was investigated so far in the pioneer works of A. Policarpo [1,2,3]. It was
shown that gas gain and electroluminescence take place simultaneously, and 
$100- 1000$ times greater light yield with respect to that of $NaI(Tl)$ 
crystal viewed in the same light-collection conditions and excellent energy 
resolution of $11.2\%$ for $5.9 \; keV$ X-rays were obtained. 
     There are the following general advantages of scintillation 
proportional counters
with respect to usual gas proportional counters.

- An energy resolution of such a counter is usually better than that 
of a gas proportional
counter with charge collection because it operates at lower values of gas gain,
and therefore, the last contributes smaller fluctuations.

- An equivalent electronic noise is $5\; -\; 10$ times lower than that of a
proportional counter with a typical preamplifier owing to the use of a PMT as
a low-noise device. This fact is very important for application of the detector
in a few-keV energy range.

- The value of HV potential is lower; thus, the breakdown problem is reduced, 
and a counter can operate at higher pressure, and therefore, possess higher 
detection efficiency for X-rays.

- The requirements on HV stability and wire uniformity is lower because the 
gas gain which is most sensitive to these parameters has lower value.

- The HV circuit is not electrically coupled with a spectrometric channel.
This eliminates possible electrical breakdown in the circuit. 

- There is no microphonic noise even for the counter of large sizes.

However, the counters of such a type are not widely used. The main reason is 
that the light-collection problem arisen for the cylindrical counter with a 
great value of the length-to-diameter ratio. One of the ways to solve this 
problem is described in [4]. Authors of this work used p-quaterphenyl as a 
wavelength shifter deposited on the inner wall of the cylinder coated with MgO 
diffusive reflector. The energy resolution of $6.4\; \%$ (FWHM) at 
$13.9\; keV$ energy of X-rays was obtained with the counter having $5\;cm$ 
diameter and $7.5\;cm$ length (the length-to-diameter ratio is 1.5). 
However, for larger ratio such a method is inefficient. \\

\section*{Method}

     Better results for large length-to-diameter ratio are possible with a 
wavelength-shifting (WLS) optical fiber readout of the electroluminescent UV 
signal. The fibers to be located near the inner surface of the cylinder. They 
reemit the UV light and thransport it through the glass window to the PMT 
placed outside the Xe volume. This method allows one to use both a PMT and 
window of a small diameter, and thus, to operate with Xe in a scintillation 
proportional mode at a pressure of several atmospheres. Moreover, large sizes 
of the counter (a length of up to 1 m and a diameter of up to 
$10 \;- \;20\; cm$) are possible without increasing PMT diameter. Optical 
fibers usually have small light attenutaion, and residual light-collection 
nonuniformity can be avoided by simultaneous recording of the light from 
opposite ends of fibers. According to [5] the yield of 0.006 photoelectrons 
per one ultraviolet photon in pure Xe can be obtained with WLS fiber 
readout. \\

\section*{Device and experimental results}

     To check this approach we have built a prototype of the counter 
schematically
shown in Fig. 1. The stainless steel tube with an inner diameter of $35\; mm$
is filled with $Xe$ under a pressure of up to $8\; atm$ . 
Xe was purified before 
filling by passing through the "Oxisorb". A total length of the 
sensitive volume 
is $\sim 20\; cm$. Eight cathode wires ($0.07 \;mm$ in diameter) are 
located at a 
distance of $\sim 1 cm$ from the central anode wire ($0.05\; mm$). The 
$~^{241}Am$ 
gamma-ray source is installed on the inner surface of the wall at a distance of 
$\sim 5\; cm$ from the end of the sensitive volume.

     Preliminary measurements of the pulse heights from the PMT coupled
with a single fiber was carried out by means of UV light source 
($\lambda =170\; nm$) placed at various distances from the PMT (Fig. 2).
A slope of the efficiency curve can be explained by self absorption of
the fiber and possible nonuniform distribution of the wavelength shifter along 
the fiber. The value of an attenuation length obtained ($L_{att} \sim 1 \;m$) 
is typical for the fiber used.

     Fig. 3 shows the pulse height spectrum measured with $~^{241}Am$
gamma-ray source at a pressure of $2 \;atm$ (see figure capture for
identification of the peaks). The photoelectron yield is about of $450
ph.e/keV$, the energy resolution is $13.3\; \%$ (FWHM) at $13.9\ keV$.

     The energy resolution of the prototype tested is worse than the
superior values achieved with scintillation proportional counters. 
Optimization of geometry and values of gas gain and electroluminescent 
amplification is necessary to improve the energy resolution of the counter.

\section*{Conclusion}

A WLS fiber readout for Xe gas scintillation proportional counter of
cylindrical geometry is proposed. This method provides possibility to make
relatively large and long counters which could be used for "dark matter"
search and neutrino magnetic moment measurement, where
massive detectors with very low energy threshold are required. Also, it 
could be used for microdosimetry and monitoring of environment.

We would like to thank A. Bolozdynya for great interest and fruitful
discussions.

\section*{References}

1.  A. Policarpo et al., Nucl. Instr. and Meth. 77 (1970) 309.

2.  A. Policarpo et al., Nucl. Instr. and Meth. 55 (1967) 105.

3.  A. Policarpo et al., Nucl. Instr. and Meth. 96 (1971) 487.

4.  H. Palmer, IEEE Trans. on Nucl. Sci. NS-22 (1975) 100.

5.  A. Parsons et al., IEEE Trans. on Nucl. Sci. NS-37 (1990) 541.

\section*{Figure caption}

\begin{description}
\item[Fig. 1]{Schematic diagram of the prototype.
1 - fibers; 2 - cathode wires; 3 - anode wire; 4 - glass window; 
5 - $~^{241}Am$.}\\

\item[Fig. 2]{Light collection efficiency of the fiber irradiated by 
170-nm UV light versus distance from PMT.}\\

\item[Fig. 3]{Pulse height spectrums obtained with $~^{241}Am$ source. 
$P = 2 atm$, $U_a = 2.9 kV$   $\sim 450 ph. e/keV$;. 1 - pedestal; 
2 - $L_{\alpha} \; \;Np \; \;(13.9 keV)$; 3 - $L_{\beta} \; \;Np \; 
\;(17.8 keV)$; 4 - $K_{\alpha} \; \;Xe \; \;(29.8 keV)$, $59.6 keV - 
K_{\alpha} \; \;Xe$; }

\end{description}

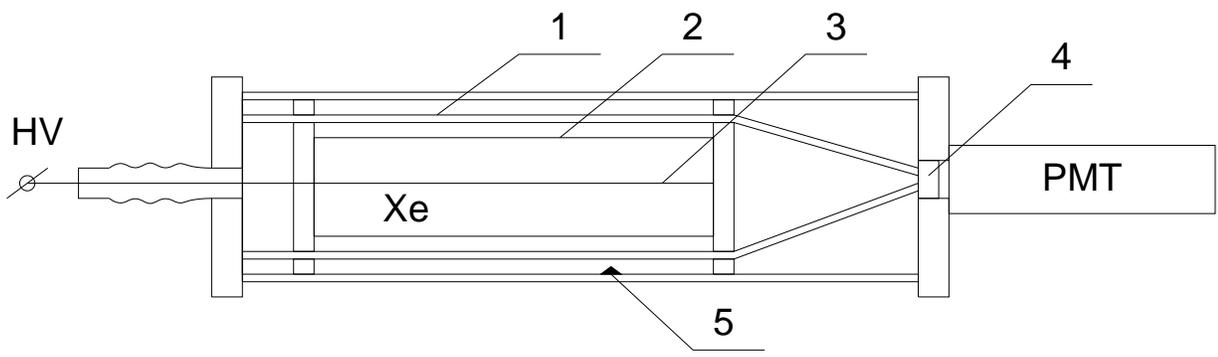
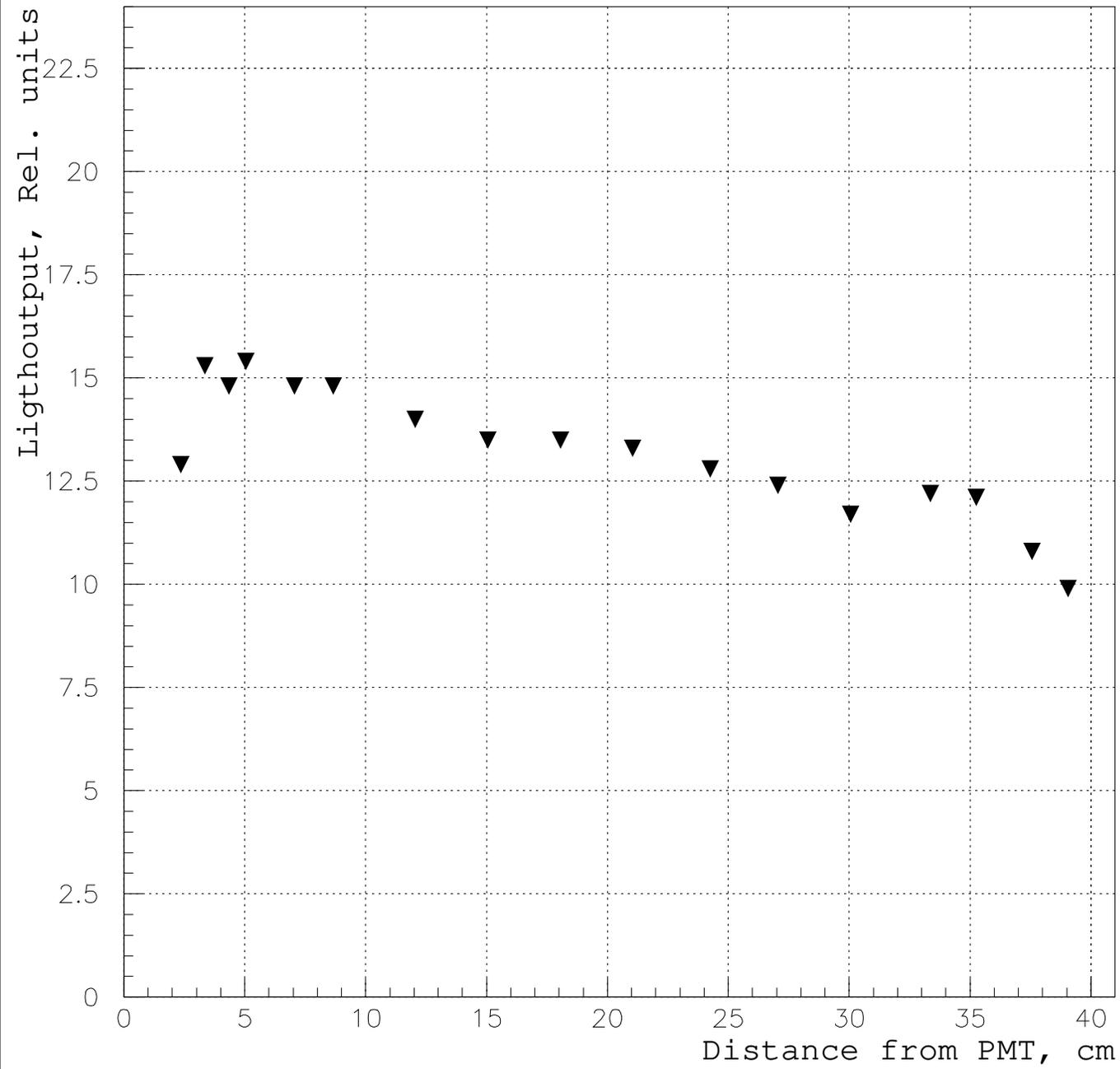
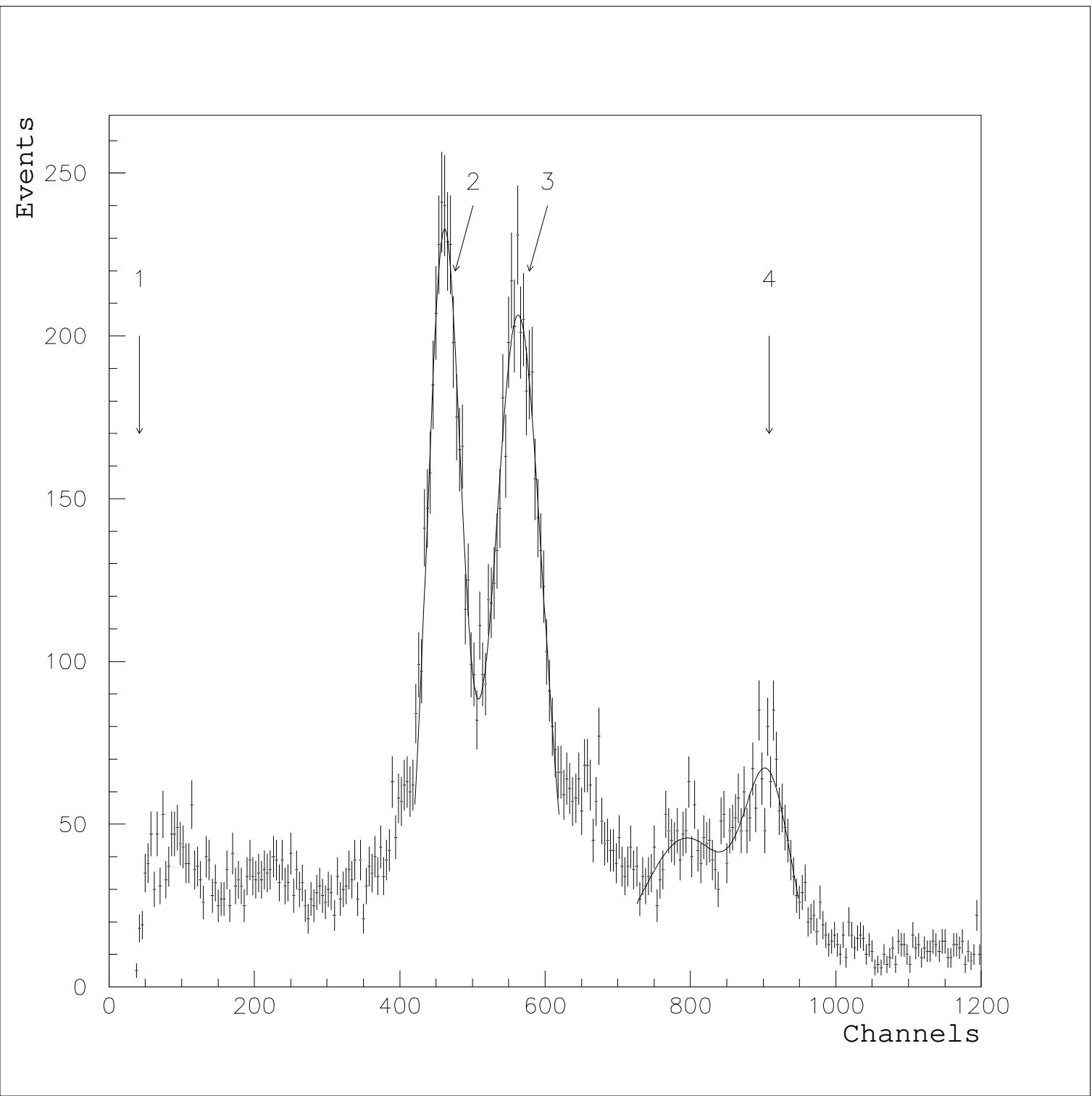
\end{document}